\documentclass[useAMS,usenatbib]{mn2e}

\usepackage{txfonts}
\usepackage{graphicx}
\usepackage{natbib}
\usepackage{amssymb}
\usepackage{epsfig}
\newcommand{\mk}{}

\title[On non-axisymmetric magnetic equilibria in stars]{On non-axisymmetric magnetic equilibria in stars}
\author[Jonathan Braithwaite]{Jonathan Braithwaite \thanks{E-mail: jon@cita.utoronto.ca} \\Canadian Institute for Theoretical Astrophysics\\60 St. George Street, Toronto ON M5S 3H8, Canada}

\pagerange{\pageref{firstpage}--\pageref{lastpage}} 
\begin{document}
\maketitle
\label{firstpage}

\begin{abstract}
In previous work stable approximately axisymmetric equilibrium configurations for magnetic stars were found by numerical simulation. Here I investigate the conditions under which more complex, non-axisymmetric configurations can form. I present numerical simulations of the formation of stable equilibria from turbulent initial conditions and demonstrate the existence of non-axisymmetric equilibria consisting of twisted flux tubes lying horizontally below the surface of the star, meandering around the star in random patterns. Whether such a non-axisymmetric equilibrium or a simple axisymmetric equilibrium forms depends on the radial profile of the strength of the initial magnetic field.
The results could explain observations of non-dipolar fields on stars such as the B0.2 main-sequence star $\tau$ Sco or the pulsar 1E 1207.4-5209. The secular evolution of these equilibria due to Ohmic and buoyancy processes is also examined.
\end{abstract}

\section{Introduction}
\label{sec:intro}


Convective stars of various types tend to have small-scale, time-varying magnetic fields, which is interpreted as dynamo action driven by differential rotation and convection/buoyancy instabilities. Stars which lack convection tend to display large-scale, steady magnetic fields.
Where we find magnetic fields in upper-main-sequence (MS) stars ($>1.5M_\odot$), which are radiative apart from a small convective core, they are large-scale and steady at least for as long as they have been observed (first detection by \citet{Babcock:1947}, see \citealt{Mathys:2001} for a recent review). Many of these stars have roughly dipolar fields, but many have rather more complex fields, such as $\tau$ Sco \citep{Donatietal:2006}. We find magnetic fields of similar geometry and total flux in some white dwarfs (see e.g. \citealt{Schmidt:2001,WicandFer:2005}; also magnetic fields in Type-Ia SN ejecta -- see e.g. \citealt{StrandSol:2007}). Again, many WDs have dipolar fields and many have stronger quadrupolar and octupolar components \citep{Beuermannetal:2007}, although none have very small-scale fields in the same way that convective stars do.

In neutron stars (NSs), there is little observational constraint on the precise geometry of the fields (see e.g. \citealt{HarandLai:2006}). However, there is evidence in some NSs for a non-dipolar configuration. For instance, the pulsar 1E 1207.4-5209 has a spindown-inferred dipole of $2 - 4\times 10^{12}$ G \citep{Pavlovetal:2002} while \citet{Sanwaletal:2002} find a surface field strength of $1.5\times 10^{14}$ G from absorption features in the spectrum. Similarly, \citet{Beckeretal:2003} find dipole and mean surface field strengths of $10^9$ and $10^{11}$ G respectively in the pulsar RBS B1821-24.

None of these non-convective stars display differential rotation. Given the lack of necessary ingredients for a dynamo, we infer that the magnetic fields are `fossil' remnants in some stable equilibrium which therefore evolve not on the relatively short Alfv\'en time-scale (years, hours and seconds in MS stars, WDs and NSs respectively) but only on longer, diffusive time-scales. The fields are stable equilibria which formed from the magnetic field left over from the last convective or otherwise `chaotic' period which the star experienced. In the case of MS stars, this is the convective/accretive protostellar phase; WDs (at least, the heavier ones which have magnetic fields) evolve mainly from the convective cores of intermediate-mass-MS stars; NSs are born out of presumably turbulent core-collapse supernovae and experience $\sim100$s of neutrino-driven convection and neutron-finger instability (see \citealt{Bonannoetal:2006}). In addition to their turbulent beginnings, two things which all these formerly convective stars have in common is the enormous range in field strengths between stars of the same class ($10^{<2-5}$, $10^{4-9}$ and $10^{11-15}$ gauss in upper-MS, WDs and NSs respectively) and the lack of correlation between the field strength and rotation speed (or rather, the inferred rotation speed at birth) which is particularly puzzling, given the clear correlation observed in {\it active} dynamos between field strength and rotation speed (see e.g. \citealt{Dibyendu:2004}).

On the theoretical side, the emphasis has been on finding magnetic field configurations which are in stable equilibrium. Analytically, one can produce an equilibrium configuration simply by making sure all forces are balanced and then test its stability to arbitrary perturbations either by finding a dispersion relation or, more often, by using an energy method \citep{Bernsteinetal:1958}. Unfortunately, it has been easier to demonstrate the instability of various configurations than to find stable configurations. For instance, both axisymmetric poloidal fields and axisymmetric toroidal fields have been shown to be unstable \citep{Wright:1973,MarandTay:1973,Tayler:1973,Braithwaite:2006a}. More recently, it has become possible to find stable equilibria using numerical methods. Braithwaite \& Nordlund (2006, hereafter Paper I; see also \citealt{BraandSpr:2004}) found such a configuration by evolving an arbitrary initial magnetic field in time and watching it find its way into a stable equilibrium. This equilibrium is roughly axisymmetric, consisting of both toroidal and poloidal components (which is logical, given that both are unstable on their own) in a twisted-torus configuration. From the outside, the field looks approximately dipolar, as do the fields of many (perhaps most) upper-MS stars and WDs. It is the main purpose of this paper to demonstrate the existence -- and explore the properties and stability -- of more complex, non-dipolar configurations.

In Section~\ref{sec:simulations} I describe numerical magnetohydrodynamic simulations where equilibria evolve from arbitrary initial conditions. I then use simple analytic methods to explore the stability and shape of these equilibria in Section~\ref{sec:analytic} and then look at the slow, quasi-static diffusive evolution of these equilibria in Section~\ref{sec:diffusion}. In Section~\ref{sec:discussion} I discuss the results in the context of astrophysical applications and observational consequences before concluding in Section~\ref{sec:conclusion}.

\section{Numerical simulations}\label{sec:simulations}

In Paper I it was found that an arbitrary initial magnetic field can evolve into a roughly axisymmetric twisted torus configuration. Here, the aim is to investigate whether that is the only stable equilibrium available or whether there are other equilibria which can be reached from different initial conditions.

\subsection{The numerical model and setup}\label{sec:numerical}

The setup of the simulations was described in some detail in Paper I; I give a brief summary here. The star is modelled as a ball of self-gravitating ideal gas (ratio of specific heats $\gamma=5/3$) of mass $M$, arranged in a polytrope of index $n=3$, so that specific entropy increases with radius and the star is stably stratified. [This has important consequences; the behaviour of a magnetic field in an isentropic star will be explored in a forthcoming paper.] The star is contained in a computational box of side $4.2R$, where $R$ is the stellar radius. Surrounding the star is an atmosphere of low electrical conductivity, which behaves like a vacuum in that the magnetic field there relaxes to a potential (current-free) field. The initial magnetic field is chosen to resemble that expected to be present at the end of a period of convection. In other words, we want to create a chaotic, small-scale field. To do this, the initial magnetic field is calculated from a vector potential (thus ensuring $\nabla\cdot{\mathbf B}=0$) which contains wavenumbers with a flat power spectrum up to a magnitude $k_{\rm max}$. The vector potential is tapered so that 
\begin{equation}
B \sim \rho^p,
\label{eq:defp}
\end{equation}
where $\rho$ is the gas density and $p$ is a free parameter. Note that if the star forms out of a uniformly magnetised cloud and the same fraction of flux is lost from all fluid elements, then we expect $p=2/3$. {\mk In all cases, some magnetic flux goes through the surface of the star; a greater amount for smaller $p$ (see Figure~\ref{fig:radprof}, where the solid lines represent initial conditions).}

The code used is the {\sc stagger code} \citep{NorandGal:1995,GudandNor:2005}, a high-order finite-difference Cartesian MHD code which uses a `hyper-diffusion' scheme, a system whereby diffusivities are scaled with the length scales present so that badly resolved structure near the Nyquist spatial frequency is damped whilst preserving well-resolved structure on longer length scales. This, and the high-order spatial interpolation and derivatives (sixth order) and time-stepping (third order) increase efficiency by giving a low effective diffusivity at modest resolution ($128^3$ here). The code includes Ohmic and well as thermal and kinetic diffusion. Using Cartesian coordinates avoids problems with singularities and simplifies the boundary conditions: periodic boundaries are used here.

As always with numerical work, one has to ensure that the relevant time-scales, which are separated in reality by many orders of magnitude, are also sufficiently separated in simulations. Here, the three relevant time-scales are the sound-crossing time, the Alfv\'en time-scale and the diffusion time-scale, with the ordering
\begin{equation}
\tau_{\rm s} \ll \tau_{\rm A} \ll \tau_{\rm d}.
\label{eq:times}
\end{equation}
The short sound crossing time ensures that the star remains in pressure equilibrium, or rather, that it evolves on an Alfv\'en time-scale in quasi-static pressure equilibrium, where I define the Alfv\'en time-scale as $\tau_{\rm A} \equiv R\sqrt{M/2E}$ where $E$ is the magnetic energy in the star. Likewise, once the magnetic forces are balanced, the star evolves on the diffusive time-scale in quasi-static MHD (and pressure) equilibrium. The strength of the magnetic field in the simulations must therefore be low enough that the Alfv\'en speed is everywhere much less than the sound speed: the ratio of thermal to initial magnetic energy is thus chosen to be $400$. However, the magnetic energy is expected to fall significantly, and to stop the Alfv\'en time-scale from becoming comparable to the diffusion time-scale, the strength of the magnetic field is artificially boosted as it decays so that the total magnetic energy stays constant. To do this, a term is added to the induction equation $\partial {\bf B}/\partial t = .... + a{\bf B}$, where $a$ is calculated at every timestep. A record is kept of the degree of boosting so that the `actual' magnetic energy can be retrieved -- it is this quantity $E'$ which is plotted in Figure~\ref{fig:enhel}, where $E'$ is calculated from $\partial E'/\partial t = -2aE'$. This holds $E$ and therefore $\tau_{\rm A}$ constant and maintains the time-scale ordering in (\ref{eq:times}).

\subsection{Formation of equilibria}\label{sec:formation}

The evolution in time of the magnetic field is followed in the simulations. The value of $k_{\rm max}$ was set to $36R^{-1}$ and various values of $p$ were tried: $0$, $1/3$, $2/3$ and $1$. In all cases, a stable equilibrium is reached after a few Alfv\'en crossing times, but both the geometry and energy of the equilibrium is found to vary. Two types of equilibrium are found: simple axisymmetric (as found in Paper I), and more complex non-axisymmetric.

\begin{figure}
\includegraphics[width=1.0\hsize,angle=0]{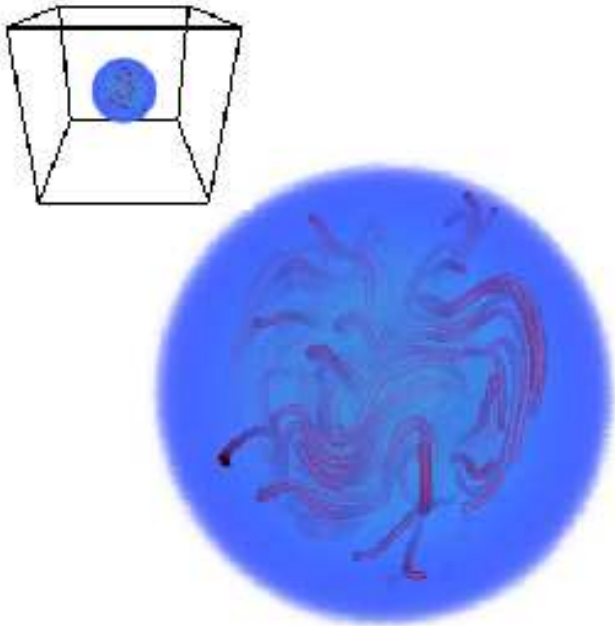}
\caption{The initial conditions: the top left shows the whole computational domain; the main picture shows just the star, which is represented by blue shading.}
\label{fig:init-conds}
\end{figure}

\begin{figure}
\includegraphics[width=1.0\hsize,angle=0]{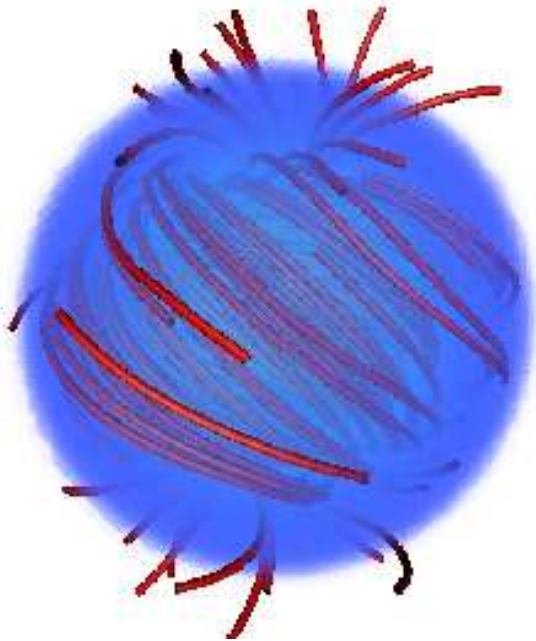}
\caption{The simplest of the equilibria, which forms when the initial magnetic field is sufficiently concentrated into the centre of the star. The configuration is roughly axisymmetric, with a twisted flux tube looped around the equator and a poloidal field passing through the middle of the star.}
\label{fig:r1}
\end{figure}

\begin{figure}
\includegraphics[width=1.0\hsize,angle=0]{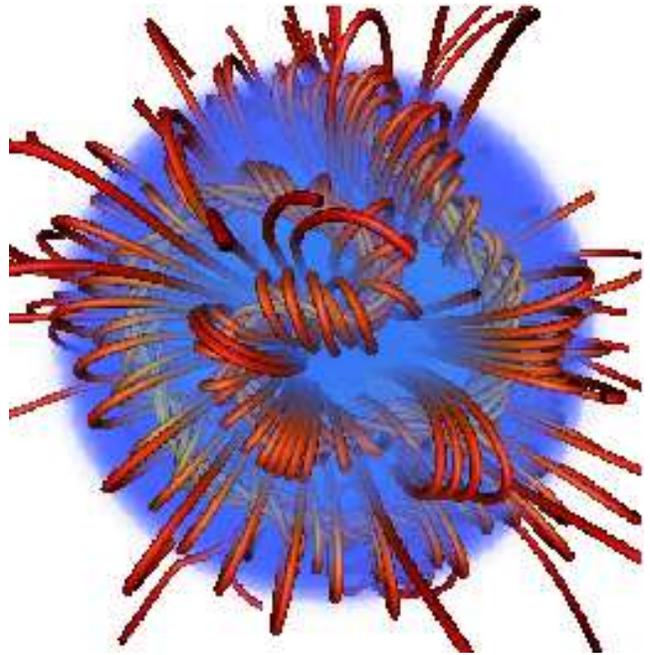}
\includegraphics[width=1.0\hsize,angle=0]{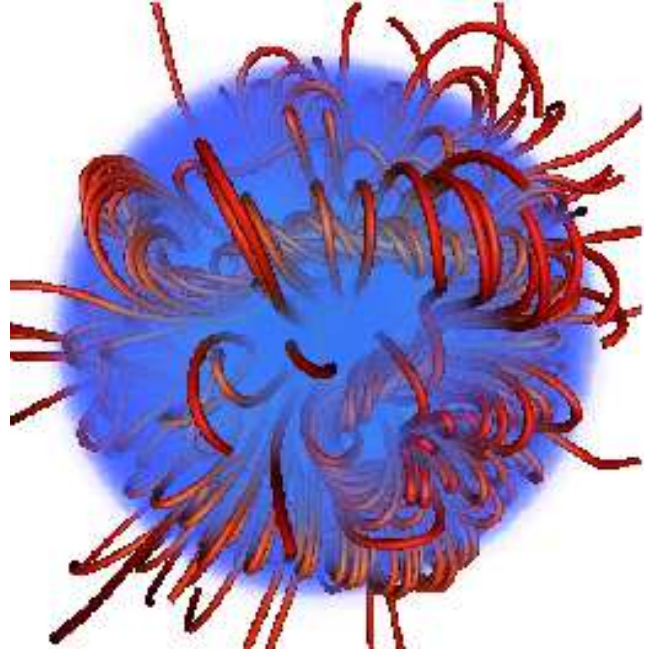}
\caption{A non-axisymmetric equilibrium viewed from opposite sides of the star. The equilibria consist of one or more twisted flux tubes buried just below the surface of the star. This kind of equilibrium forms if the initial magnetic field is not concentrated into the centre of the star. In making this figure, the `opacity' of the star (blue shading) was set so that the field lines just below the surface can be seen but that lines further inside cannot, as a visual aid.}
\label{fig:r0}
\end{figure}

Figure~\ref{fig:init-conds} shows the initial conditions (of the $p=1$ run). Note that the star takes up only a small part of the box -- this wastes some computational power but is necessary to ensure an approximately correct treatment of the potential field outside the star. For the run with $p=1$, the magnetic field after around $9$ Alfv\'en crossing times is displayed in Figure~\ref{fig:r1}. The configuration is roughly axisymmetric and contains both toroidal and poloidal components.

When $p=0$, a non-axisymmetric equilibrium is reached. This is displayed in Figure~\ref{fig:r0}; the figure shows the star from both sides. The configuration consists of twisted flux tubes below the surface of the star. The flux tubes lie horizontally and meander around the star at some distance below its surface. 

\section{Stability of non-axisymmetric fields}
\label{sec:analytic}

In this section, the structure and stability properties of the non-axisymmetric field will be examined, and predictions made about how the nature of the equilibrium depends on the conserved parameters energy, helicity and flux.

\subsection{The structure of a twisted flux tube}
\label{sec:structure}

\begin{figure}
\includegraphics[width=1.0\hsize,angle=0]{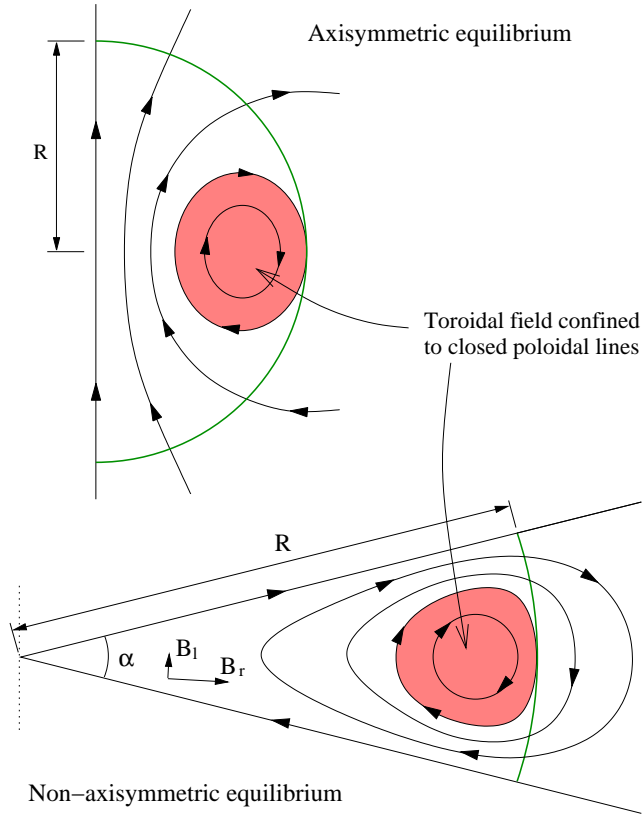}
\caption{{\mk Cross-sections of a twisted flux tube below the surface of a star of radius $R$. Above, the axisymmetric case where the flux tube is a circle around the equator; below, the non-axisymmetric case where the flux tube meanders around the star in some more complex fashion.} The directions of $B_{\rm l}$ and $B_r$ are indicated by the arrows in the lower part of the figure; $B_{\rm t}$ is directed into the page. The poloidal field lines are marked with arrows; it can be seen that the boundaries of the wedge are magnetic surfaces. The toroidal field (direction into/out of the paper, shaded area) is confined to the poloidal lines which are closed within the star.}
\label{fig:x-sec}
\end{figure}

Consider a twisted flux tube of length $s$ and width (at the surface of the star) $\alpha R$ whose axis lies at some constant radius inside a star of radius $R$, {\mk so that $\alpha$ is the angle subtended by the flux tube at the centre of the star}. The cross-section of such a flux tube is illustrated in the lower part of Figure~\ref{fig:x-sec}. It is assumed that $\alpha \ll1$ (see section~\ref{sec:energy}). The three components of the magnetic field are toroidal (parallel to the length of the flux tube), radial and latitudinal; average values are denoted respectively by $B_{\rm t}$, $B_{\rm l}$ and $B_r$. The radial and latitudinal components will be collectively referred to as the poloidal component. {\mk [Note that this coordinate system is local to each position along the flux tube.]} Since the flux tube cannot have any ends, it must be joined in a loop. This loop can be wrapped around the star in any fashion; it is assumed below that the radius of curvature is {\mk significantly} greater than the tube's width. 

\begin{figure}
\includegraphics[width=1.0\hsize,angle=0]{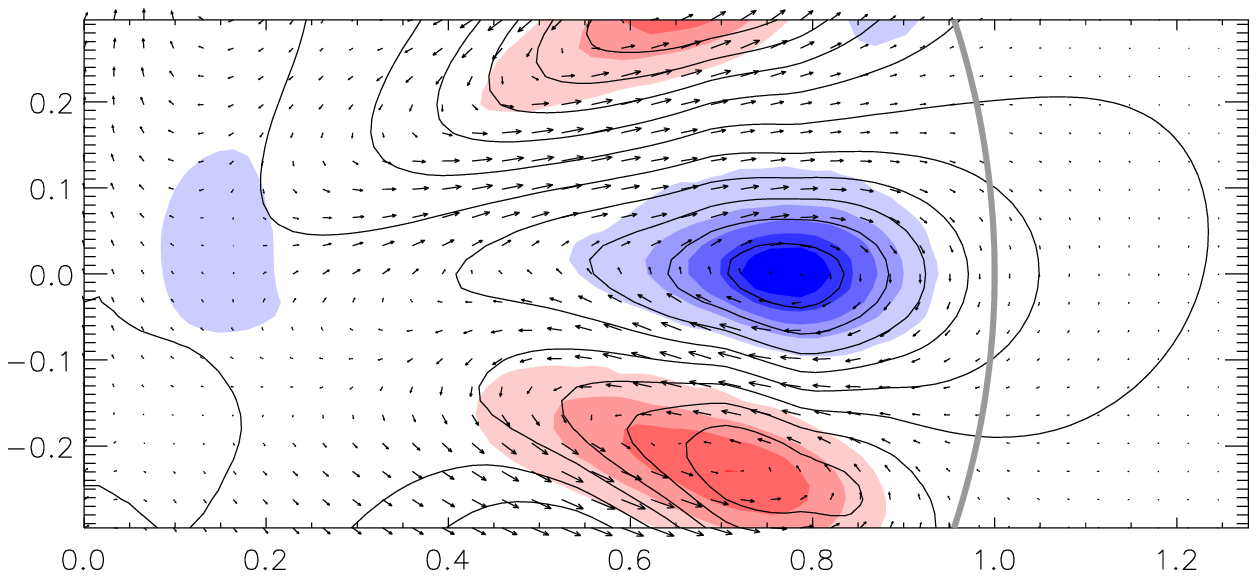}
\includegraphics[width=1.0\hsize,angle=0]{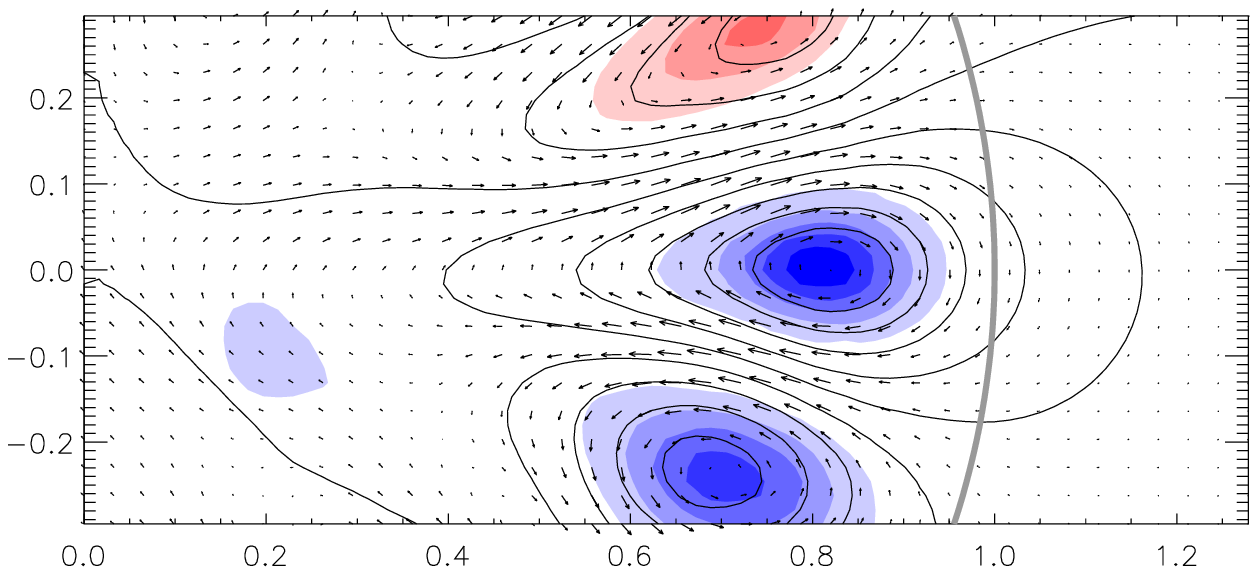}
\includegraphics[width=1.0\hsize,angle=0]{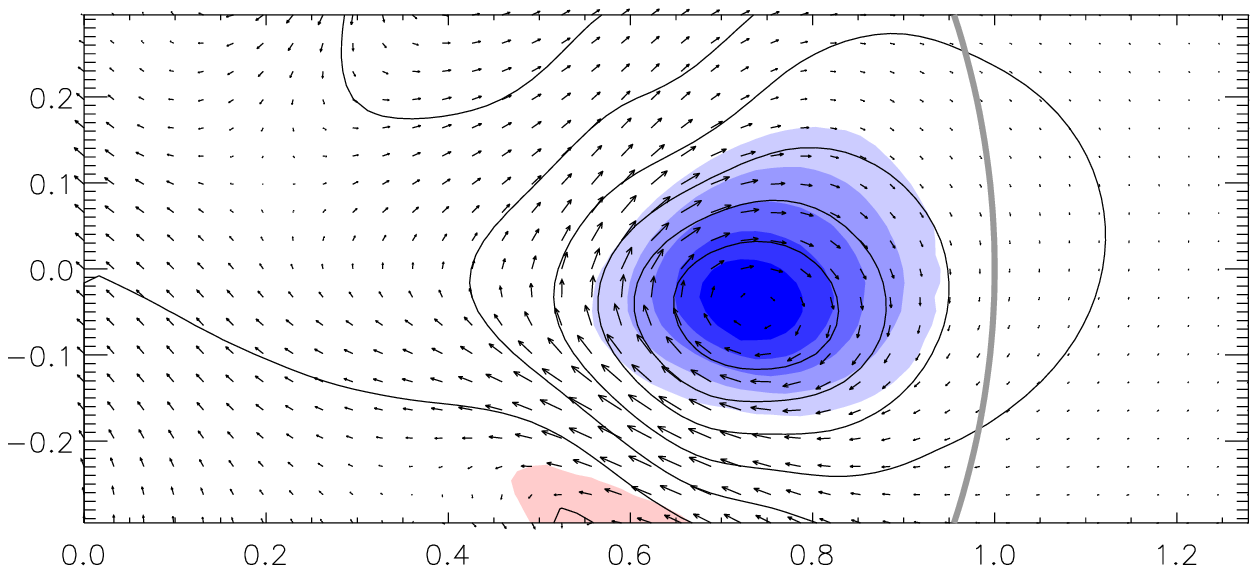}
\includegraphics[width=1.0\hsize,angle=0]{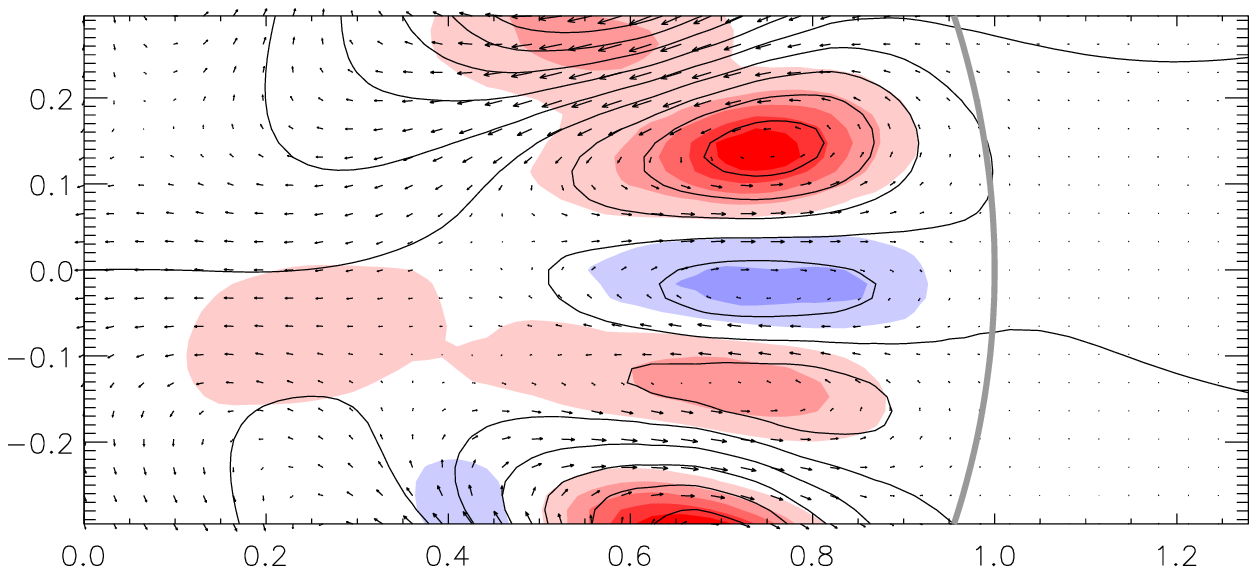}
\includegraphics[width=1.0\hsize,angle=0]{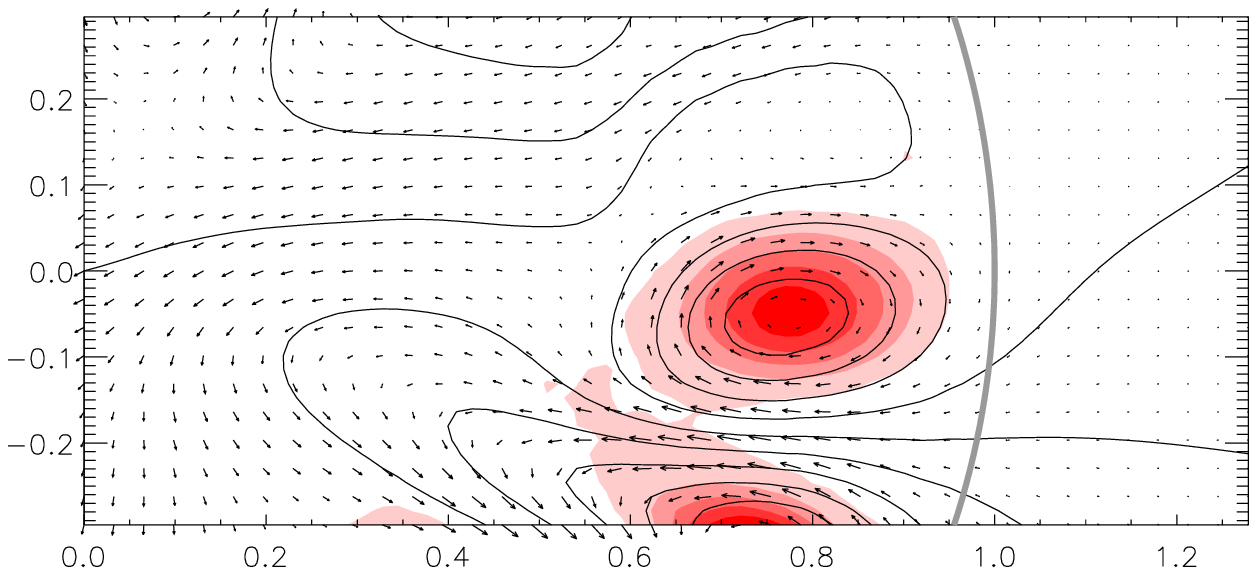}
\caption{Cross-sections of flux tubes in the run with $p=0$ at time $t=17\tau_{\rm A}$. The curved grey line towards the right is the surface of the star and the centre of the star is on the left. The blue/red shading represents the toroidal field component (out of/into the page) multiplied by $\varpi$, the horizontal coordinate in these cross-sections, {\mk the cylindrical radius.} The poloidal component is represented by the arrows and by contours of its scalar potential. It can be seen that the arrows are very nearly parallel to the contours of the scalar potential, showing that the length scale of variation in the direction perpendicular to the page is much greater than the length scale in the plane of the page, i.e. that the flux tubes meander around the star over scales much greater than their width.}
\label{fig:sim-x-secs}
\end{figure}

We can immediately compare this picture to the simulations. In Figure~\ref{fig:sim-x-secs} we have five flux-tube cross-sections from the run with $p=0$ at time $t=17\tau_{\rm A}$. The upper three frames are centred on different parts of one flux tube; the lower two frames on different flux tubes. {\mk In each case, the axis of the flux tube is perpendicular to the page.} Toroidal field is represented by blue/red shading and poloidal field is represented by arrows as well as by contours of the scalar potential $\psi$ {\mk (see e.g. equation 3.6 of \citealt{MarandTay:1973})} defined as
\begin{equation}
\frac{\partial\psi}{\partial z} = -\varpi B_\varpi \;\;\;\; {\rm and} \;\;\;\;  \frac{\partial\psi}{\partial\varpi} = \varpi B_z. 
\end{equation}
{\mk Here, it is more correct to use cylindrical coordinates $\varpi$ and $z$, as opposed to the spherical $r$ and $\rm{l}$ above, although the two sets are the same in the limit of small $\alpha$. The axis of the flux tube lies on (and in fact defines the location of) both the $z$ and $\rm{l}$ axes, so that both these coordinate systems are local, as opposed to the usual cylindrical system for instance where $z$ has the same direction at every location.}
Now, this scalar potential is only meaningful if the divergence of the poloidal field is negligible, i.e. that the relevant length scale along the flux tube (i.e. its radius of curvature) is much greater than the tube's width. That the arrows in Figure~\ref{fig:sim-x-secs} (representing the real direction of the poloidal field) and contours of $\psi$ are roughly parallel confirms that this is a reasonable approximation.

Mestel (1961, see also \citealt{Roxburgh:1966}) found that in a star with an axisymmetric field, the condition that the azimuthal component of the Lorentz force vanishes (necessary since it cannot be balanced by the pressure gradient) gives the condition
\begin{equation}
{\bf B}_{\rm p}\cdot\nabla(\varpi B_{\rm t}) = 0,
\end{equation}
where ${\mathbf B}_{\mathrm p}$ is the poloidal field, $B_{\rm t}$ is the azimuthal (toroidal) component and $\varpi$ is the cylindrical radius. In other words, poloidal field lines are contours of $\varpi B_{\rm t}$. In the non-axisymmetric case the same argument applies: $\varpi$ in this context is the distance from the line passing through the centre of the star, perpendicular both to the flux tube axis and to a line connecting the centre of the flux tube to the centre of the star (i.e. the dotted line on the left of Figure~\ref{fig:x-sec}). It is also shown in Figure~\ref{fig:x-sec} that the toroidal component of the field is confined to the volume enclosed by the largest poloidal line closed within the star, and is zero outside this volume. {\mk The reason for this is that long-lasting currents cannot exist outside the star and consequently the toroidal field outside the star has to be zero, and that $\varpi B_\phi$ has to have the same value along a given poloidal line. A hand-waving argument as to how the toroidal field outside the largest closed poloidal line is destroyed is as follows. A} poloidal line which crosses the stellar surface would `unwind' if it contained any toroidal component. In this process, toroidal field is redistributed along the whole of the poloidal line. Toroidal field in the atmosphere is then destroyed by reconnection on a short time-scale, and the toroidal field inside the star is redistributed along the poloidal line to where it has been lost. This occurs on the Alfv\'en time-scale. In this way, all toroidal field on the poloidal line is quickly transferred into and destroyed in the atmosphere. {\mk By analogy, one can think of twisted elastic bands: those which are connected up in a loop stay wound, while those with free ends unwind.}

These two properties are consistent with the results from the simulations described in the previous section -- in Figure~\ref{fig:sim-x-secs} we see that the poloidal field lines do roughly coincide with contours of $\varpi B_{\rm t}$, and that the toroidal field is very weak outside of the last closed poloidal line. In the figure, the contours are evenly (linearly) spaced, and weak $B_{\rm t}$ of either direction is coloured white.

\subsection{Reaching equilibrium: energy considerations}
\label{sec:energy}

{\mk The volume of the flux tube is $s\alpha R^2/3$ and the} magnetic energy it contains is
\begin{equation}
E = \int{\frac{B^2}{8 \pi} dV} \approx \frac{s \alpha R^2}{3} \frac{1}{8 \pi} \left[ B^2_{\rm t} + B^2_{\rm l} + \left(1 + \frac{3 a^2_r \alpha}{2 \pi}\right)B^2_r \right],
\label{eq:energy}
\end{equation}
where $s$ is the length of the tube (the length on the surface of the star), $\alpha$ is the angle subtended at the centre of the star and $R$ is the radius of the star. {\mk The first two terms in the square brackets and the first term in curved brackets represent the field energy inside the star from each of the three components, and the last term represents the field energy outside the star. The field outside the star is potential (current-free) so can be calculated by solving the Laplace equation, given $B_r$ at the surface. The form of $B_r$ at the surface is not known, so I have simply assumed a basic sinusoidal form with the coefficient $a_{\rm r}$ (of order, but less than unity) relating the mean radial field component inside the flux tube to the radial field component on the surface. It will be seen below that the energy above the surface is not important unless $\alpha$ is very small.}

We now allow the the length and width of the flux tube to change adiabatically. The flux tube will stretch or contract until $\partial E/\partial \alpha=0$, i.e. an equilibrium is reached\footnote{It is straightforward to verify that $\partial^2 E/\partial \alpha^2 > 0$ at this point, i.e. that the equilibrium is stable to stretching/contracting perturbations.}. {\mk Now, we are considering a star where the magnetic pressure is everywhere very much less than the thermal, so magnetic forces can have only a negligible effect on the gas pressure. Also, the star is stably stratified so that energy is required to move fluid elements in the radial direction since work has to be done against the buoyant restoring force, and again, the magnetic forces are weak in comparison to the buoyancy force. We therefore}
have the following restrictions on the displacements $\mathbf{\xi}$ which take place during this adjustment to equilibrium:
\begin{equation}
{\bf \nabla}\cdot{\bf \xi} \approx 0 \;\;\;{\rm and}\;\;\; {\bf r}\cdot{\bf \xi} \approx 0.
\label{eq:zerodiv}
\end{equation}
{\mk In other words, weak magnetic forces have to avoid doing work against stronger forces. These approximations are standard in analysis of MHD instabilities in stars (e.g. \citealt{PitandTay:1986}; \citealt{Spruit:2002}).} The consequence of these restrictions is that as long as only reversible processes are allowed, the surface-field factor $a_r$ stays constant as does the volume of the flux tube, since no matter can leave or enter it, which can be expressed by stating that the area of the flux tube on the stellar surface $F = s \alpha R = \mathrm{const}$. {\mk The radial distribution of magnetic energy also cannot change since gas and therefore the magnetic energy frozen into it cannot be transported in the radial direction.} In addition, we have:
\begin{equation}
\frac{\partial \ln B_{\rm t}}{\partial \ln \alpha} = -1 \:,\;\;\;\;\frac{\partial \ln B_{\rm l}}{\partial \ln \alpha} = 1 \;\;\;\;{\rm and}\;\;\;\; \frac{\partial \ln B_r}{\partial \ln \alpha} =  0,
\label{eq:ddalpha}
\end{equation}
which follow simply from flux-freezing as well as from (\ref{eq:zerodiv}).
We can now differentiate (\ref{eq:energy}) with respect to $\alpha$ and set $\partial E/\partial \alpha=0$, giving
\begin{equation}
B^2_{\rm t} \approx B^2_{\rm l} + \frac{3 a_r^2 \alpha}{4\pi} B^2_r \approx
B^2_{\rm l}\left(1 + \frac{3 a_r^2}{4\pi\alpha}\right),
\label{eq:deda}
\end{equation}
{\mk where we have noted from the geometry that $B_{\mathrm l} \approx \alpha B_r$ to arrive at the part on the right hand side. Remembering that $a_r$ is somewhat less than $1$, we now look at the case where $\alpha$ is large enough so that the term in (\ref{eq:deda}) containing it can be ignored. We are left with}
\begin{equation}
B_{\mathrm t} \approx B_{\mathrm l} \approx \alpha B_r.
\label{eq:BtBl}
\end{equation}
Therefore the adjustment to equilibrium consists in stretching or contracting until the toroidal and latitudinal components are roughly equal. This principle is also (partially) applicable to axisymmetric equilibria -- see Section \ref{sec:ratios}.

{\mk On the other hand, if $\alpha$ is very small, (\ref{eq:BtBl}) becomes instead
\begin{equation}
\frac{4\pi\alpha}{3 a_r^2}B_{\mathrm t} \approx B_{\mathrm l} \approx \alpha B_r.
\label{eq:BtBlalt}
\end{equation}
However, at least in Ap stars there is no observational evidence for this, i.e. for magnetic fields with very small-scale structure. In any case, a small-scale field would evolve and decay on a timescale shorter than the stellar lifetime. (\ref{eq:BtBl}) will be used below instead of (\ref{eq:BtBlalt}), although it should be noted that this point might conceivably need to be revisited in future.}

\subsection{Magnetic flux and helicity}
\label{sec:helicity}

If we define a vector potential ${\mathbf A}$ from ${\mathbf B}=\nabla\times{\mathbf A}$, it can be shown \citep{Woltjer:1958} that the magnetic helicity $H\equiv\int{{\mathbf B}\cdot{\mathbf A} dV}$, {\mk which is a global rather than local quantity,} is conserved in the limit of infinite conductivity\footnote{Magnetic helicity is gauge-independent only if the boundaries of the region of integration are either periodic or magnetic surfaces with ${\mathbf B}\cdot d{\mathbf S} = 0$. The boundaries of the flux tubes here (Figure~\ref{fig:x-sec}) are indeed magnetic surfaces; note that ${\mathbf B}$ goes to zero at infinity.}.

{\mk Consideration of the length scales $R$, $\alpha R$ and $R$ in the toroidal, latitudinal and radial directions respectively and} from the definition of ${\mathbf A}$ we see that
\begin{equation}
B_{\rm t} \sim \frac{A_r}{\alpha R} - \frac{A_{\rm l}}{R}\:,\;\;\;\;B_{\rm l} \sim \frac{A_{\rm t}}{R}\;\;\;\;{\rm and}\;\;\;\;B_r \sim \frac{A_{\rm t}}{\alpha R},
\label{eq:hel}
\end{equation}
{\mk which using (\ref{eq:BtBl}) becomes $A_{\rm t} \sim A_{\rm l} \sim A_r/\alpha \sim RB_{\rm t}$}. The helicity of the flux tube is now given by
\begin{equation}
H \equiv \int{{\bf B}\cdot{\bf A} \,dV} \sim 2 F B_r \Phi_{\rm t} \approx 4 \Phi_{\rm p} \Phi_{\rm t}
\end{equation}
where $\Phi_{\rm t} \approx (1/2)\alpha R^2 B_{\rm t}$ and $\Phi_{\rm p} \approx (1/2)s\alpha R B_r = (1/2) F B_r$ are the toroidal and poloidal fluxes; note that all coefficients of order unity are rather approximate.
The helicity, as well as both $\Phi_{\rm t}$ and $\Phi_{\rm p}$, are conserved during adiabatic changes in $\alpha$ and $s$ {\mk and evolve only due to diffusive processes; this is described in Section~\ref{sec:diffusion}. It will be seen in Fig~\ref{fig:enhel} that helicity is indeed roughly conserved during the formation of equilibria on the Alfv\'en timescale, so remains a useful tool to aid our understanding of dynamical relaxation into equilibrium.}

Using (\ref{eq:BtBl}), we can now give $\alpha$ as a function of the conserved quantities defined above:
\begin{equation}
\alpha^2 \approx \frac{2}{R^2} \frac{\Phi_{\rm t}}{B_r} \approx \frac{F}{R^2} \frac{\Phi_{\rm t}}{\Phi_{\rm p}},
\label{eq:alpha2}
\end{equation}
so that if we know the area $F$ and the ratio of toroidal to poloidal fluxes of a tube, we can calculate its equilibrium width $\alpha R$ and length $s$.

The above arguments can be applied not only to the flux tube as a whole, but also to any small length $\delta s$ of the tube with area $\delta F$. Along the length of the flux tube $\Phi_{\rm t}$ must be constant, but $B_r \approx 2 \delta \Phi_{\rm p}/\delta F$ can vary. There is no reason why a single flux tube needs to have a uniform width -- at every point along its length, there is a local equilibrium giving $\alpha$ as a function of the toroidal flux and the local value of $B_r$ (remember from (\ref{eq:ddalpha}) that $B_r$ is constant during adiabatic changes in $\alpha$). This can be seen in the first three plates in Figure~\ref{fig:sim-x-secs}, which are centred on the same flux tube at different points along its length -- at the position of the third plate, $\alpha$ is greater than at the other two.

Neighbouring flux tubes need to have their poloidal components in opposite senses in order to avoid a discontinuity at the boundary between them. If the neighbouring tubes are actually parts of the same tube, then the toroidal components will clearly also be in opposite senses, i.e. in the cross-section, if the toroidal field of the tube goes into the page, the toroidal field of the neighbouring tube section will be directed out of the page. If the neighbours are part of other tubes, the poloidal components still need to have opposite senses but there is no constraint on the direction of the toroidal part. In fact, there is no way that one flux tube can `know' about the direction of the toroidal field in any of the other tubes present, because they are insulated from one another by regions of zero toroidal field. In the first plate of Figure~\ref{fig:sim-x-secs} we see that the two tube sections above and below the tube in the centre have both poloidal and toroidal fields in opposite senses to the tube in the centre, so that all three have helicity (which is essentially just the product of the toroidal and poloidal fluxes) of the same sign, and may in fact be different parts of the same tube. In the second plate, the lower tube has opposite helicity to the other two tube sections visible. This means then that the equilibrium has lower net helicity than an otherwise identical configuration with the toroidal field of this tube reversed; if we also reduced the strength of this new configuration we could make a field of equal helicity but lower energy.

\begin{figure}
\centerline{\hbox{\epsfxsize=2.7in
   \epsffile{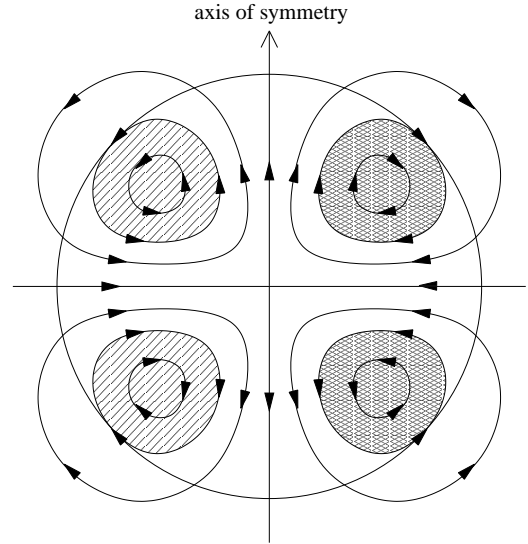}}}
\caption{The cross-section of a star containing the simplest imaginable zero-helicity equilibrium. The shaded areas represent the toroidal component of the field -- dark (light) shading represents field going into (out of) the page. The vertical axis is the star's axis of symmetry.}
\label{fig:zero-hel}
\end{figure}

{\mk Therefore, there are many stable equilibria with the same helicity but different energies, and it is possible that an arbitrary initial field will evolve on the Alfv\'en timescale, whilst conserving helicity, into a higher-energy, local equilibrium rather than into the lowest energy state for its helicity.} Stable configurations with {\it zero} net helicity are possible in principle. The cross-section of a simple zero-helicity field is illustrated in Figure~\ref{fig:zero-hel}. {\mk Whether such a higher-energy local equilibrium can actually be reached from realistic initial conditions is another question. Also, it is plausible that diffusive evolution (where helicity is no longer conserved) on a much longer timescale may coax the configuration into a lower-energy equilibrium.}

\subsection{Simulations: a quantitative analysis}

\begin{figure}
\includegraphics[width=1.0\hsize,angle=0]{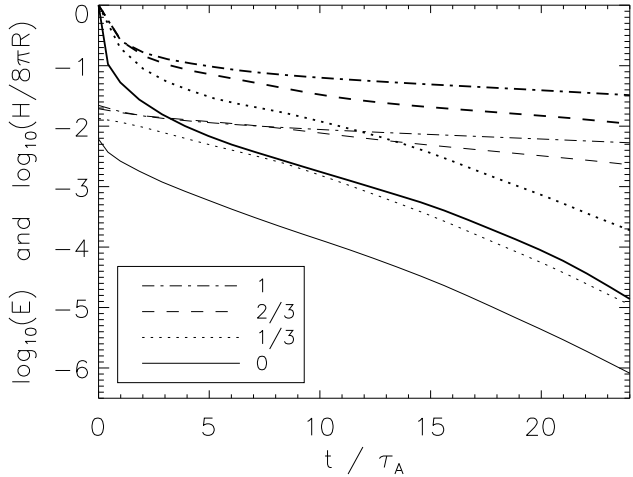}
\caption{Log magnetic energy $E$ and helicity $H/8\pi R$ (thick and thin lines respectively) against time, for runs with various values of the central-concentration parameter $p$. Helicity has been divided by $8\pi R$ to give it units of energy and so that the ratio of the two is equal to the helicity length as a fraction of the stellar radius, $r_{\rm H}/R$.}
\label{fig:enhel}
\end{figure}

It is now informative to have a more quantitative look at the simulations. First, one can measure the magnetic helicity and magnetic energy. In Figure~\ref{fig:enhel}, these are plotted for the runs with $p=0$, $1/3$, $2/3$ and $1$ ($p$ was defined in (\ref{eq:defp})). Time, on the horizontal axis, is given in terms of Alfv\'en times $\tau_{\rm A}\equiv R\sqrt{M/2E}$ where $E$ is the magnetic energy. All runs have the following in common: the helicity falls throughout at some low rate, while the energy falls significantly during the first couple of Alfv\'en time-scales before settling down to a similar rate of decay as the helicity. This can be interpreted as the formation of a stable equilibrium, followed by decay of the equilibrium configuration on a longer, diffusive time-scale. Differences between the runs are also visible in the figure: once equilibrium has been reached, the runs with $p=1$ and $2/3$ experience a slower drop in energy and helicity than the runs with lower values of $p$. This is presumably because the length scales in the axisymmetric configurations which form in the $p=1$ and $2/3$ runs are greater and the diffusive time-scales correspondingly greater.

\begin{figure}
\includegraphics[width=1.0\hsize,angle=0]{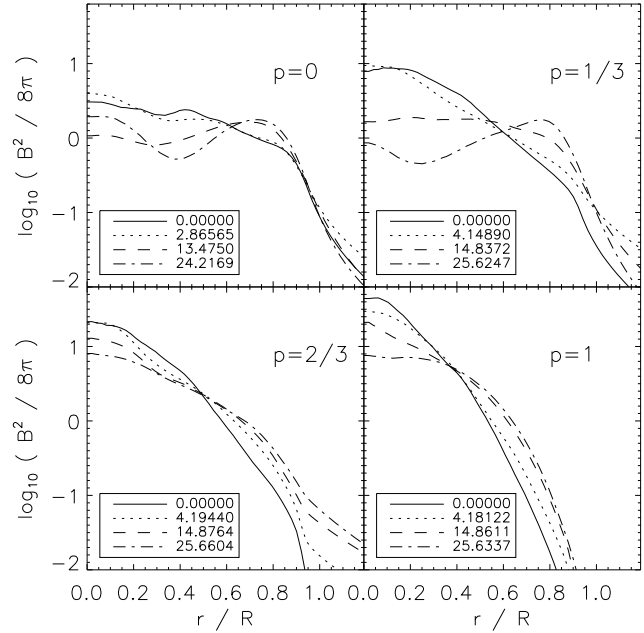}
\caption{Radial profiles of magnetic energy density $B^2/8\pi$ (log, as a fraction of the mean) at various times for the four runs with different values of $p$. In each plot the lines correspond to the times (in Alfv\'en time-scales) indicated in the boxes. It can be seen that the profiles change little during the formation of the equilibria, and change only on the longer, diffusive time-scale.}
\label{fig:radprof}
\end{figure}

We can also look at the radial profile of the magnetic energy density $B^2/8\pi$ to see where most of the energy resides. In Figure~\ref{fig:radprof} are plotted radial profiles of the magnetic energy density at different points in time for each of the four runs with different values of $p$. We can see that the energy profiles change little over the Alfv\'en time-scale when the equilibrium is forming, but that on a longer time-scale it becomes shallower and eventually the energy density peaks at around $r\approx 0.8R$, where the axes of the flux tubes are located. Transport of flux in the radial direction is hindered by the stable stratification and can only take place on a diffusive time-scale. Precisely which kind of diffusive process (magnetic or thermal) is responsible is discussed in Section~\ref{sec:diffusion}.

\begin{figure}
\includegraphics[width=1.0\hsize,angle=0]{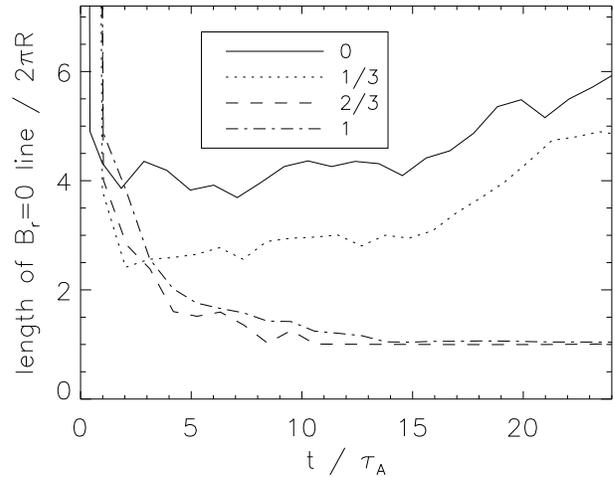}
\caption{The length of the $B_r=0$ lines on the surface of the star, divided by the circumference of the star (denoted by $W$), against time, for runs with various values of the central-concentration parameter $p$: $0$, $1/3$, $2/3$ and $1$. For the higher values (centrally-concentrated initial field) a dipolar field is formed, while for the lower values of $p$ a more complex equilibrium results.}
\label{fig:wiggliness}
\end{figure}

It is also useful to measure the length of the flux tubes. The easiest method is to measure the length of the lines on the surface where $B_r=0$, as the flux-tube axes always lie directly below these lines. Figure~\ref{fig:wiggliness} is a plot of $W$, the total of the lengths of all flux tubes in terms of the stellar circumference, against time, for the runs with $p=0$, $1/3$, $2/3$ and $1$. Clearly, the runs with higher $p$ values, which form axisymmetric equilibria, should have values of $\approx 1$ because the flux tube is essentially a great circle around the star\footnote{In fact, what I refer to here as `axisymmetric' is only approximately so. The equilibria are generally slightly offset from the centre of the star and may be a little warped in shape. This is presumably a result of the random nature of the initial conditions.}. The non-axisymmetric equilibria have higher values, because the flux tubes are longer and meander around the star in a more complex fashion; amongst these equilibria, a lower value of $p$ leads to longer, narrower flux tubes. If we assume that the whole volume of the star is taken up with flux tubes of similar width then $\alpha \approx 2/W$ in the limit of small $\alpha$.

\begin{figure}
\includegraphics[width=1.0\hsize,angle=0]{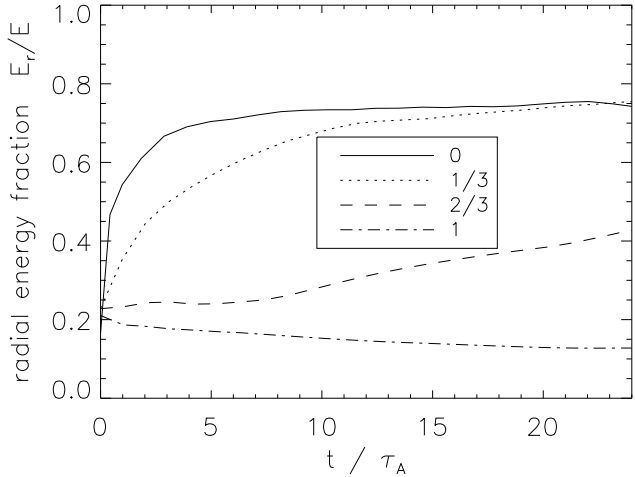}
\caption{The energy in the radial component of the field $B_r$ as a fraction of the total, against time for runs with various values of the central-concentration parameter $p$: $0$, $1/3$, $2/3$ and $1$. As expected, the non-axisymmetric equilibria have greater radial energy fractions.}
\label{fig:merome}
\end{figure}

The energy in the radial component of the field as a fraction of the total energy, $E_r/E$, is plotted in Figure~\ref{fig:merome}. In light of (\ref{eq:BtBl}) we expect that $E_r/E \approx 1/(1+2\alpha^2)$, although (\ref{eq:BtBl}) is really only valid if the radial energy profile is fairly flat. If we look for instance at the $p=0$ run at $t=17\tau_{\rm A}$ (when the cross-sections in Figure~\ref{fig:sim-x-secs} were taken), we see that from Figure~\ref{fig:wiggliness} that $W\approx 4.5$, so that $\alpha\approx 0.44$ and we expect $E_r/E\approx 0.72$, which corresponds approximately to the value of $\approx 0.75$ in Figure~\ref{fig:merome}.

\section{Diffusive evolution of equilibria}
\label{sec:diffusion}

First, let us assume that the electrical conductivity is high enough and that thermal conductivity is low enough that the relevant diffusive time-scales are much larger than the Alfv\'en time-scale, i.e. that $(\alpha R)^2/\eta \gg \alpha R / v_{\mathrm A}$ and $(\alpha R)^2/\kappa \gg \alpha R / v_{\mathrm A}$, where $\eta$ and $\kappa$ are the magnetic and thermal diffusivities respectively and $v_{\mathrm A}$ is the Alfv\'en speed. Then the flux tube remains in quasi-static equilibrium described by (\ref{eq:BtBl}) and (\ref{eq:alpha2}) while it is evolving diffusively. This is the case in the simulations, as we can see from Figure \ref{fig:enhel} that the equilibrium evolves diffusively over many Alfv\'en time-scales. Furthermore, assume that the area $F$ of the flux tube remains constant. This is justified if the flux tube(s) have filled up all of the available space on the surface of the star, i.e. that $F=4\pi R^2$; the simulations show this to be roughly correct, although there are some small `empty' spaces, one of which is visible in the last plate of Figure~\ref{fig:sim-x-secs}.

In the simulations, we see that the total magnetic energy in the equilibria falls, and that the energy moves to larger radii (Figures~\ref{fig:enhel} and \ref{fig:radprof}), both taking place on some diffusive time-scale (rather than on the Alfv\'en time-scale on which the equilibrium forms). The overall fall in energy (and helicity, the product of toroidal and poloidal fluxes) can be understood as an obvious consequence of finite conductivity, which can also explain the {\it flattening} of the radial energy profile; what is more difficult to explain is the evolution from a flat profile into a state where the energy density is higher at larger radii than in the centre of the star. Nor can finite conductivity easily explain why $\alpha$ should fall -- in Figure~\ref{fig:wiggliness} we see that the length of the flux tubes is gradually increasing, and $\alpha$ is therefore falling. From (\ref{eq:alpha2}) we can infer from a falling $\alpha$ that the toroidal flux $\Phi_{\rm t}$ is falling faster than the poloidal flux $\Phi_{\rm p}$. These phenomena can perhaps be better understood in terms of thermal diffusion.

Nevertheless let us first look at the effect of Ohmic dissipation on the flux tubes. Finite conductivity causes toroidal field to leak out of the volume enclosed by the largest closed poloidal field line, onto poloidal lines which go through the surface of the star. Via the mechanism described in Section \ref{sec:structure}, this toroidal field is quickly destroyed. The effect of finite conductivity on the {\it poloidal} component is that gradients in the radial and latitudinal directions are smeared, causing the overall amplitude to go down. Also, since the direction of the poloidal component is changing over shorter length scales (in the horizontal direction) deeper in the star, this could explain why the field strength deeper inside the star drops faster than that further out. Also, perhaps since the toroidal field is confined to a smaller space than the poloidal, it should decay faster since the diffusion time-scale is shorter for smaller length scales, thus explaining the fall in time of $\alpha$.

\subsection{Buoyancy}

A more convincing explanation for the rise of flux tubes to the surface and for the greater loss of toroidal than poloidal flux involves {\it thermal} rather than magnetic diffusion. [There is an analogous process, in the case where the stratification is due not to an entropy gradient but to a composition gradient and the relevant diffusivity is that of chemical elements rather than heat. Here, I shall illustrate the principle only for the case of entropy stratification and thermal diffusion.] This process can cause the field to rise towards the surface of a stably stratified star in the following way. Imagine a magnetised region surrounded by non-magnetised plasma. It must be in pressure equilibrium with the surroundings, so that the sum of its magnetic and thermal pressures is equal to the thermal pressure outside. If it has the same temperature as the surroundings, it will be less dense, and rise. In a stability stratified star (specific entropy increasing with radius) the magnetised region will cool adiabatically as it rises, its temperature falling faster than the ambient temperature, and it will eventually reach an equilibrium where its total pressure and density are equal to those of its new surroundings. However, it will now have a lower temperature than the surroundings, and any thermal diffusion will cause it to absorb heat and rise further into a new equilibrium. Now, this mechanism should cause our flux tubes to rise and even in the absence of magnetic diffusion inside the star, could lead to destruction of the toroidal flux as poloidal lines breach the surface of the star and their toroidal component is lost to reconnection processes in the atmosphere.

Consider a flux tube of radius $a$, containing toroidal (axial) and poloidal field components $B_{\rm t}$ and $B_{\rm p}$ surrounded by an unmagnetised gas of pressure $P$ and temperature $T$. The tube is thin, so $a\ll H_P$ where $H_P$ is the pressure scale height. The pressure exerted by the field on the surrounding gas, which is equal to the difference in external and internal thermal pressures, is given by
\begin{equation}
\Delta P = -\frac{1}{2\pi a}\frac{\partial}{\partial a}\left(\pi a^2 \frac{B_{\rm t}^2+B_{\rm p}^2}{8\pi}\right) = \frac{B_{\rm t}^2}{8\pi}
\end{equation}
where the expression in brackets represents the magnetic energy per unit length of the tube. Note that the poloidal component does not contribute, which comes from the fact that adiabatic expansion of the tube does not change the energy of that component. Denoting quantities internal and external to the tube by i and e, we have at all times
\begin{equation}
\rho_{\rm e} = \rho_{\rm i} \;\;\;\;\; {\rm and} \;\;\;\;\; P_{\rm e} = P_{\rm i} + \Delta P.
\end{equation}
Assuming an ideal gas equation of state gives us $\Delta T/T = \Delta P/P$, and the heat flow into the tube per unit length is
\begin{equation}
Q \approx 2\pi a \;\kappa \rho \frac{R_\mu \Delta T}{a},
\end{equation}
where $R_\mu = R/\mu$ is the gas constant. After some algebra, the vertical speed $v$ of the flux tube is found to be
\begin{equation}
\frac{v}{H_P} \equiv \tau_{\rm rise}^{-1} \approx \frac{2\kappa}{a^2 \beta_{\rm t}} \left(\frac{\nabla_{\rm a}}{\nabla_{\rm a}-\nabla}\right)
\label{eq:rise}\end{equation}
where $\beta_{\rm t}\equiv8\pi P/B_{\rm t}^2 \gg 1$ and
\begin{equation}
\nabla\equiv \frac{{\rm d}\ln T}{{\rm d}\ln P} \;\;\;\;\; {\rm and} \;\;\;\;\; \nabla_{\rm a}\equiv {\left(\frac{\partial \ln T}{\partial \ln P}\right)}_s
\end{equation}
are derivatives of the surroundings and of the equation of state at constant entropy, respectively. Note that the speed becomes infinite as the temperature gradient approaches the adiabatic gradient where convection sets in; in reality the speed would be limited by factors not considered here such as aerodynamic forces. Except near the boundaries of a stably-stratified zone, we can treat the expression in brackets in (\ref{eq:rise}) as a factor of order unity.

Strictly speaking, the flux tubes in these non-axisymmetric magnetic equilibria are not surrounded by a non-magnetised medium, although there are gaps between the toroidal-field regions of neighbouring flux tubes. Nor is $a\ll H_P$; however, it seems likely that expression (\ref{eq:rise}) is still applicable here, albeit only rather approximately. More quantitative, local analysis and simulations (perhaps also in 2-D) could shed light on this process and on the speed at which the flux tubes rise. Very approximately, we have
\begin{equation}
\tau_{\rm rise} \sim (\alpha R)^2 \beta / \kappa \sim \alpha^2 \beta \tau_{\rm KH}
\label{eq:risetime}
\end{equation}
where $\beta$ is the ratio of thermal to magnetic energy and $\tau_{\rm KH}$ is the Kelvin-Helmholtz (thermal) time-scale. Whether this is longer or shorter than the Ohmic diffusion time-scale depends on the type of star under consideration (see Section \ref{sec:comparison}); in the simulations presented here, thermal and magnetic diffusivities are equal and $\beta$ ranges from $\sim10^4$ in the centre of the star to $\sim30$ near the surface.

\subsection{Horizontal movements}

As the length and width of the flux tubes adjust quasi-statically to the loss of toroidal flux, their arrangement under the surface of the star can be expected to change, with tubes pushing each other around in the horizontal direction. These movements will be slow for the most part, but fast readjustments may also occur, particularly if stress builds up between tubes and a different arrangement of flux tubes becomes energetically accessible. The observational consequences of this are not very clear; in addition to a change in the geometry of the magnetic field one might expect some release of energy accompanied by an apparent perturbation to the rotational phase (and also to the rotation period, if the moment of inertia is changed).

\subsection{The depth and cross-sectional shape of the flux tubes}\label{sec:depth}

From (\ref{eq:alpha2}) we can predict the width of a flux tube every point along its length if we know its toroidal flux and the local value of $B_r$. However, we still have no argument to explain the cross-sectional shape of the tube, or rather, the shape of the part where the toroidal field resides. Even in the same simulation at the same time-step, we see a variety of shapes, ranging from circular to very elongated in the radial direction (see Figure~\ref{fig:sim-x-secs}), whilst the depth of the tubes does appear fairly uniform. It looks therefore that the radial distribution of toroidal flux, and very likely the radial distribution of energy, are globally fairly uniform and that only the flux-tube width $\alpha$ varies from place to place. This uniformity could be a result of the stable stratification and the restrictive effect it has on the movement of fluid and therefore magnetic flux in the radial direction, in contrast to movements in the horizontal direction caused by changes in the width $R\alpha$ of the flux tubes. We therefore expect the radial distribution of energy in the newly-formed equilibria to be similar to that in the initial conditions.

Once the initial equilibrium has formed, the configuration continues to evolve on a slower, diffusive time-scale as discussed in the previous section, and as this happens, the radial distribution of energy changes (Figure~\ref{fig:radprof}). After some time, when the flux tubes have risen sufficiently and the magnetic energy density is low in the central part of the star, we can define a depth $fR$ above which most of the energy is contained, and redefine $B_{\rm t}$, $B_{\rm l}$ and $B_r$ as the average values in that volume within $fR$ of the stellar surface. (\ref{eq:BtBl}) becomes
\begin{equation}
B_{\mathrm t} \approx B_{\mathrm l} \approx \frac{\alpha}{f} B_r,
\label{eq:BtBlnew}
\end{equation}
since the energy argument in Section~\ref{sec:energy} is unchanged, but the different geometry changes the relation between the two poloidal components. Equations (\ref{eq:hel})-(\ref{eq:alpha2}) are changed only by factors of order unity. If we now let $f$ change while keeping $\alpha$ and $s$ fixed, we find that
\begin{equation}
\frac{\partial \ln B_{\rm t}}{\partial \ln f} = -1 \:,\;\;\;\;\frac{\partial \ln B_{\rm l}}{\partial \ln f} = -1 \;\;\;\;{\rm and}\;\;\;\; \frac{\partial \ln B_r}{\partial \ln f} =  0
\label{eq:derivs2}
\end{equation}
and that the equilibrium condition $\partial E/\partial f=0$ yields
\begin{equation}
B^2_{\rm t} \approx B^2_r - B^2_{\rm l}.
\end{equation}
Combining with (\ref{eq:BtBlnew}) we have $f \approx \sqrt{2} \alpha$ and
\begin{equation}
B_{\rm t} \approx B_{\rm l} \approx \sqrt{2} B_r.
\end{equation}
Clearly the depth $f$ of the flux tube cannot change adiabatically on an Alfv\'en time-scale as $\alpha$ can, but one at least expects that diffusive processes will tend to produce flux tubes of roughly circular or somewhat radially-elongated cross-section. This process can also be understood in another way. In general, a circular flux tube can be force-free if the hoop stress of the poloidal field is balanced by the force resulting from the gradient in the toroidal field. If the tube is now squashed into a tube of elliptical cross-section whilst maintaining its cross-sectional area, the hoop stress of the poloidal component is still equal on the major and minor axes. However, the gradient of toroidal field is stronger now along the minor axis than along the major, resulting in a restoring force pushing the tube back into its original shape. A flux tube can only be force-free if it has circular cross-section, corresponding to the lowest energy state for a given helicity.

Looking at the simulations (Figure~\ref{fig:sim-x-secs}) we see that most of the flux tubes do indeed appear to have $f$ similar to or somewhat greater than $\alpha$. However the fourth frame of the figure has $f$ significantly greater than $\alpha$. This probably comes from effects not considered here such as pressure from neighbouring flux tubes or details of the buoyancy mechanism which brings the tubes upwards.

\subsection{Axisymmetric equilibria}\label{sec:diff-axi}

In Paper I we found that an axisymmetric equilibrium gradually moves outwards, the torus expanding. The rate at which the field on the surface increases was found -- in the case where the field is {\it deeply buried} inside the star -- to be proportional to the magnetic diffusivity, which works simply to smooth out any gradients in field strength so that any centrally-concentrated field becomes more spread out, and therefore become stronger nearer the surface although the total energy of the field is falling\footnote{This is equivalent to metal bar which is initially hot in the centre: heat is lost through the sides, but is also conducted along the bar so that at first the ends become warmer despite a fall in the total thermal energy.}. However, in Paper I we did not look at the effect of the two types of diffusivity when the axis of the torus is closer to the surface and the radial energy profile is flatter. It seems likely that here, {\it magnetic} diffusion alone cannot explain the evolution and that thermal diffusion is important. 

In an axisymmetric equilibrium, the buoyant-rising effect of thermal diffusion is resisted by the tension of the toroidal component. As \citet{Reisenegger:2007} points out, this is likely to be more important when the torus is buried deep inside the star, which could explain why magnetic diffusion may be more important in deeply buried fields\footnote{He also notes that in a star with a barotropic equation of state or which is otherwise non-stably stratified, a flux tube should rise to the surface on a dynamical time-scale. The issue of stable fields in non-stably-stratified stars will be looked at in a forthcoming paper.}.

\begin{figure}
\includegraphics[width=1.0\hsize,angle=0]{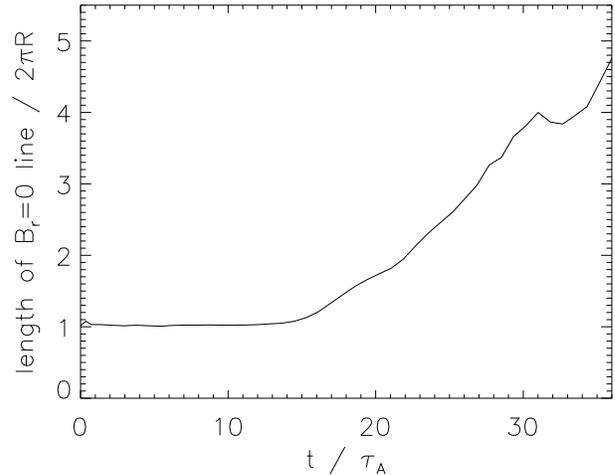}
\caption{The length of the $B_r=0$ line on the stellar surface (divided by the circumference) for an initially axisymmetric equilibrium which crosses this threshold between axisymmetric and non-axisymmetric equilibria due to toroidal flux loss.}
\label{fig:axi-non-axi}
\end{figure}

\begin{figure*}
\includegraphics[width=0.33\hsize,angle=0]{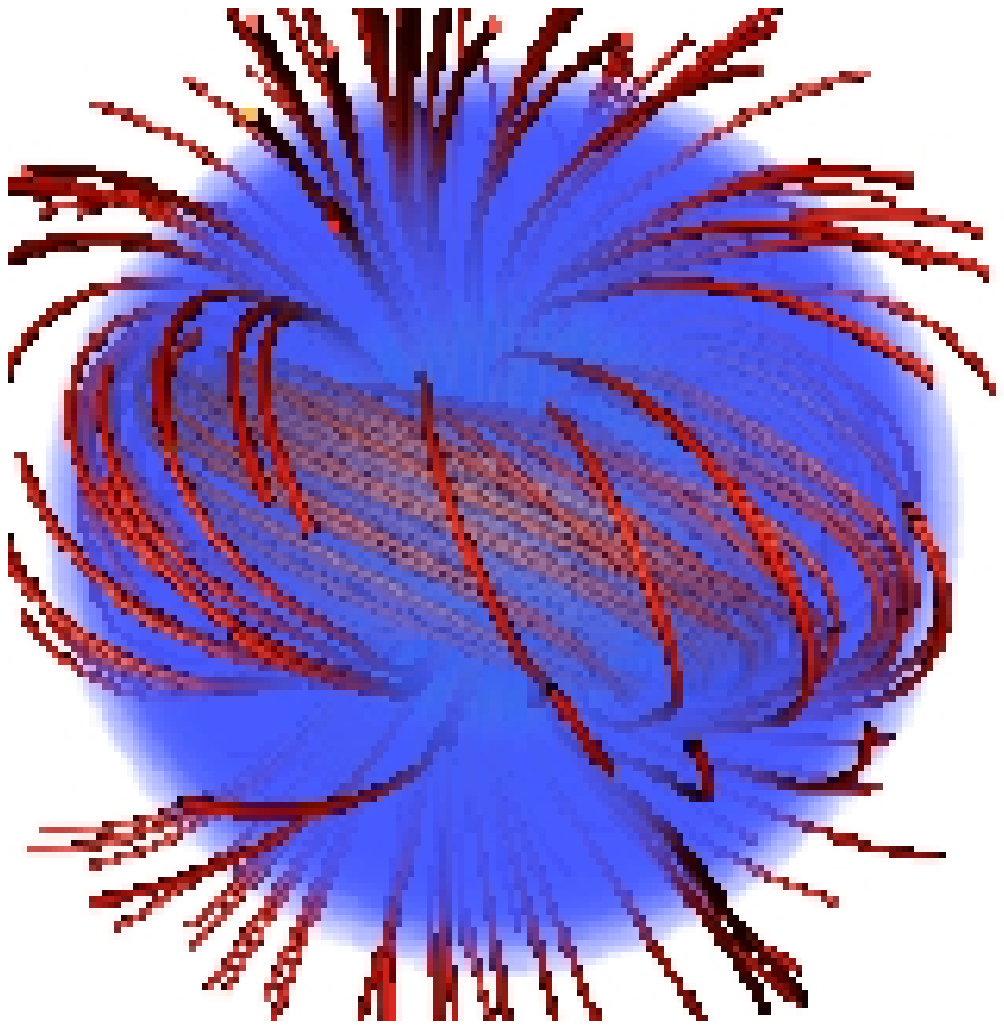}
\includegraphics[width=0.33\hsize,angle=0]{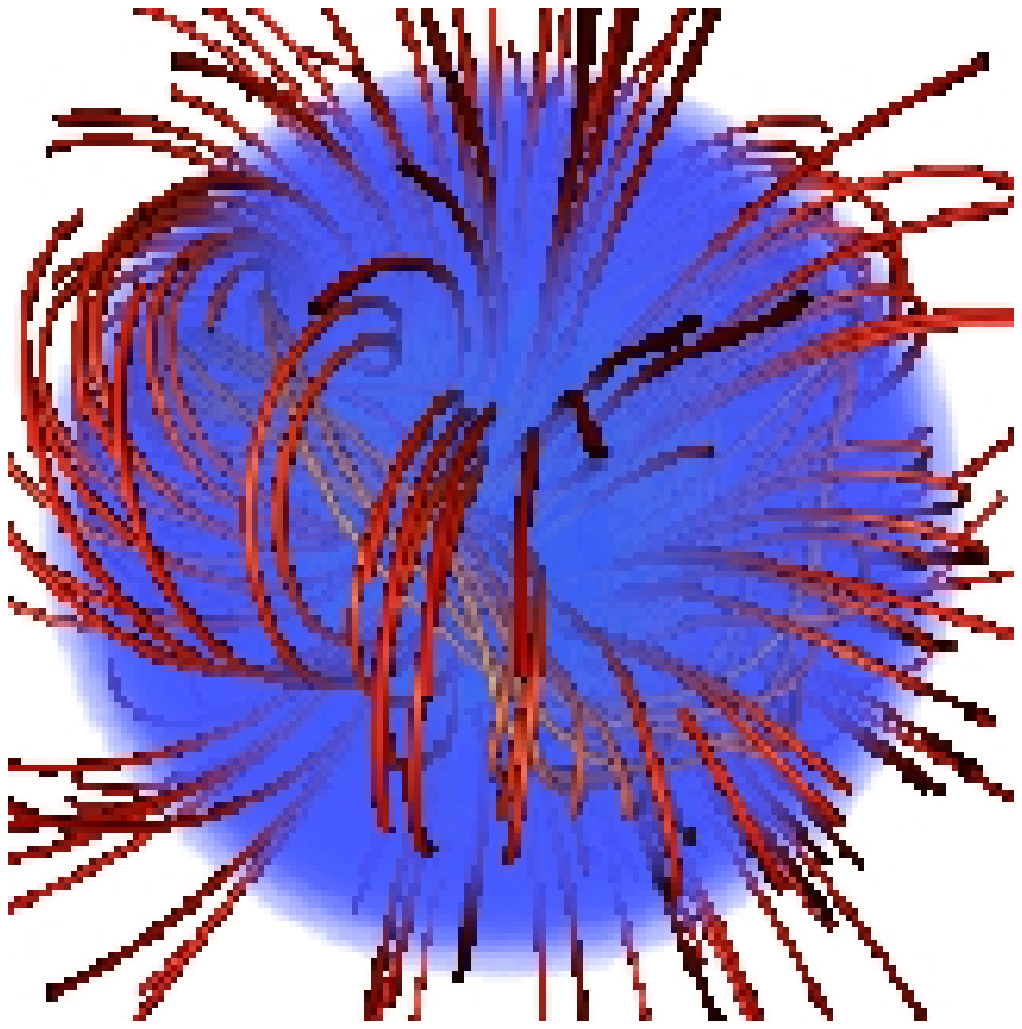}
\includegraphics[width=0.33\hsize,angle=0]{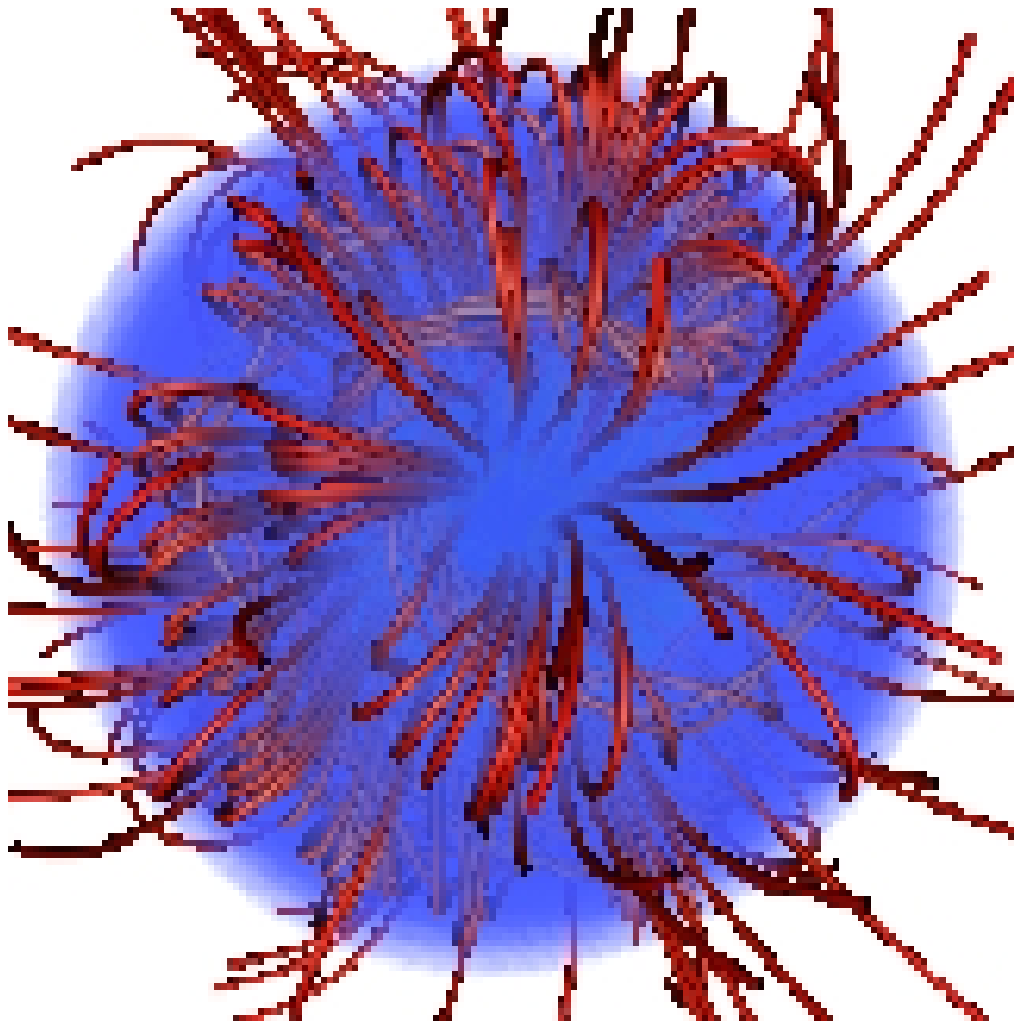}
\caption{The successive snapshots of an initially axisymmetric equilibrium which has evolved from that in Figure~\ref{fig:r1}. A transition is made from axisymmetric to non-axisymmetric configuration as diffusive processes destroy toroidal flux and it becomes energetically favourable for the circular flux tube to lengthen and become narrower. The resulting equilibrium is equivalent to that reached directly from initial conditions with a flatter radial energy profile (Figure~\ref{fig:r0}).}
\label{fig:snake}
\end{figure*}

As axisymmetric equilibrium with a flatter radial energy profile diffuses outwards, it experiences the same loss of toroidal flux as the non-axisymmetric equilibria. A {\it purely} poloidal axisymmetric field is unstable around the neutral line, i.e. where the poloidal field goes to zero. The instability, which drives poloidal field loops in a direction parallel to the field's axis of symmetry, has recently been seen in simulations (\citealt{BraandSpr:2006}, and in rotating stars in \citealt{Braithwaite:2007}). \citet{MarandTay:1973,MarandTay:1974} and \citet{Wright:1973} find that the instability can be suppressed by addition of a toroidal field of at least $\sim1/4$ of the strength of the poloidal component. This works because the instability increases the length of the neutral line, so it has to work against the tension of the toroidal component. The stronger the toroidal field, the smaller azimuthal wavenumbers are stabilised and just below the threshold, only the largest-scale ($m=2$) azimuthal mode is unstable. This is because the larger wavenumbers stretch the toroidal field more than the smaller wavenumbers, so are easier to stabilise. Therefore as an initially stable mixed poloidal-toroidal field diffuses outwards and loses toroidal flux, it eventually becomes unstable to this $m=2$ mode. Until now, the non-linear development of this instability had not been studied. In light of the arguments in Section~\ref{sec:energy}, it can be seen that once this threshold is crossed, it becomes energetically favourable for the flux tube to become longer and narrower. In other words, $\alpha$ falls from $180^\circ$ to a smaller value, and the equilibrium becomes non-axisymmetric; in fact, exactly of the type described above. Since the analysis given in Section~\ref{sec:energy} assumed small $\alpha$, it is not immediately obvious whether this transition takes place quasi-statically or whether $\alpha$ falls quickly from $180^\circ$ to an equilibrium at some lower value. We can appeal here to simulations, which appear to show a gradual transition on a diffusive time-scale. Figure~\ref{fig:axi-non-axi} shows $W$, the length of the $B_r=0$ line on the stellar surface (divided by the circumference) for an initially axisymmetric equilibrium which crosses this threshold due to toroidal flux loss. The rise in $W$ does not seem sudden, but happens only on a diffusive time-scale. Figure~\ref{fig:snake} shows the field at three times: before, just after, and some time after this transition, corresponding to $t/\tau_{\rm A}=2.9$, $16.1$ and $23.5$ in Figure~\ref{fig:axi-non-axi}.

In addition, the instability of an axisymmetric poloidal field presents us with another way of understanding the adjustment, in a non-axisymmetric configuration, of a narrower flux tube to equilibrium described in Section~\ref{sec:energy}. If the toroidal component in the flux tube is a little too weak, there is a long-wavelength instability around its axis, the resultant latitudinal displacements increasing the length of the flux tube and taking energy from, and weakening, the poloidal component until stability is restored. Therefore as the toroidal flux is gradually lost into the atmosphere, the flux tube adjusts quasi-statically and is always marginally unstable.

\section{Discussion}\label{sec:discussion}

In this section I consider some applications of these results to real stars.

\subsection{Comparison to observations}\label{sec:comparison}

As outlined in Section~\ref{sec:intro}, both axisymmetric and more complex equilibria occur in all three classes of non-convective star: upper-MS, WDs and NSs. This implies either that the initial conditions vary -- in particular the degree of central concentration of the initial field and/or flux connection through the star's surface, or that all stars form axisymmetric equilibria which then evolve after some time into more complex configurations via the process described in Section~\ref{sec:diff-axi}. Another way of looking at this is that there is a continuous series of equilibria and that diffusive processes move the star along this series. At the beginning of the series are deeply-buried axisymmetric fields, which evolve on the magnetic diffusion time-scale $R^2/\eta$, where $\eta$ is the magnetic diffusivity. As this happens, the field strength on the surface may increase at first and then decrease. After that are less deeply-buried axisymmetric fields with a greater fraction of their flux going through the star's surface; these evolve on a time-scale of either $R^2/\eta$ or $R^2\beta/\kappa$ where $\kappa$ is the thermal diffusivity; the field strength on the surface is likely to fall with time. After that come the non-axisymmetric fields which evolve on a time-scale of $\alpha^2R^2\beta/\kappa$ or $\alpha^2R^2/\eta$; since $\alpha$ is continuously falling, so the time-scale of evolution falls until the field evaporates completely. During this phase, the field strength on the surface is likely to fall on a magnetic diffusion time-scale, but the dipole field strength could fall more quickly. The star begins its life somewhere on this sequence and moves along with time. It is useful now to look at the actual time-scales in each of the classes of stably-stratified star.

In an intermediate-mass MS star, the Alfv\'en time-scale is $\sim10$ years for a field of $1$kG and the magnetic and thermal diffusion time-scales are $R^2/\eta\sim10^{10}$ years and $R^2/\kappa\sim10^6$ years; the lifetime of the star is typically $10^{8 - 9}$ years. Composition gradients in the radiative envelope are probably unimportant and since $\beta\sim10^{6-12}$, the buoyant rise time-scale {\mk -- as given by (\ref{eq:risetime}) --} is greater than the lifetime of the star, unless the flux tubes are very narrow. {\mk Very narrow flux tubes would be found observationally by a very strong average field modulus compared to the average line-of-sight component, which is not seen (see e.g. \citealt{Mathys:2001}). There should not therefore be any} significant evolution of the field during the main sequence.  Indeed, there is arguably no observational evidence for field evolution during the main-sequence (but see \citealt{Hubrigetal:2000a}). We need therefore different initial conditions. It is possible that stars are born with a range in central concentrations around the threshold so that some stars develop axisymmetric fields and some more complex equilibria, but still with large $\alpha$. The questions of the origin of the range in field strengths and the apparent cutoff at $\sim 200$ gauss (no stars are observed with fields weaker than this {\mk down to an observational limit of a few gauss} -- see \citealt{Auriereetal:2007}) are still open, but since the former is generic to all non-convective stars and the latter is unique to A stars, it is likely that the solutions to these two questions are unconnected, and that there is some mechanism in A stars to destroy fields below a certain threshold. {\mk Whether there is a similar cutoff in O and early B stars as that in A and late B stars remains to be seen.}

In WDs, the Alfv\'en, magnetic and thermal diffusion time-scales are a few days (field of $10$ MG), $\sim10^{10}$ and $\sim10^4$ years respectively \citep{ChaandGab:1972}; Hall drift may also be important \citep{Muslimovetal:1995}. Because the thermal diffusivity is so high, the thermal buoyancy timescale given by (\ref{eq:risetime} is less than the cooling time ($10^{9-10}$ yr) for the stronger fields. However, a composition gradient can prevent buoyant rise because the species diffusion in WDs is very low. Since carbon and oxygen have almost exactly the same mass per electron, C-O WDs lack composition stratification but the heavier O-Ne WDs are stratified. This could be the reason that we observe that the magnetic WDs tend to be heavier than the non-magnetic WDs {\mk \citep{WicandFer:2005}}. Observationally, no correlations of magnetic properties with age have been found. As in A stars, a mix of dipole-dominated and quadrupolar and higher order fields are found, but none dominated by very high order harmonics.

In NSs, the Alfv\'en time-scale is $\sim100$s for a field of $10^{12}$ G. The diffusive time-scales are subject to considerable uncertainly from both the theory -- because Ohmic diffusion, ambipolar diffusion and Hall drift as well as superconductivity and superfluidity may all play roles -- and from the observations. There is evidence that the fields of radio pulsars decay on time-scales of $>10^6$ years while magnetar fields seem to decay faster, perhaps over around $10^4$ years {\mk (see e.g. \citealt{WooandTho:2004}).} Hall drift and buoyant rise provide natural explanations for this as they scale with the field strength, as $B$ and $B^2$ respectively. Buoyancy of flux tubes in neutron stars depends mainly on the composition gradient, the difference with WDs being that not only can chemical elements diffuse into or out of a flux tube but that the chemical elements can reach equilibrium via the neutrino-emitting reaction $n \leftrightarrow p + e$ {\mk \citep{Reisenegger:2007}.} The speed of the latter does not depend on the length scale, so fields with narrow flux tubes would evolve no faster than simpler fields if this were the dominant mechanism. Some NSs are observed with rather small-scale fields, such as 1E 1207.4-5209 whose dipole field is $\approx 50$ times smaller than the typical field strength on the surface {\mk (\citealt{Pavlovetal:2002}, \citealt{Sanwaletal:2002}).} Such a field could be a low-$\alpha$ equilibrium. This kind of equilibrium could provide an explanation for the observed phenomenon of NSs with similar dipole field strengths and rotation periods which have quite different observational properties, some being observed as radio pulsars and some as magnetars \citep{Morrisetal:2002}. Magnetars with $10^{14}$G dipole fields could have a complex fields containing much more energy than the radio pulsars with the same dipole field strength, which could account for the high energy output of magnetars.

\subsection{Permissible poloidal/toroidal ratios in axisymmetric fields}\label{sec:ratios}

In an axisymmetric configuration, it is only the poloidal component we can see on the surface and which manifests itself in various processes taking place outside the star, such as disc accretion, torques in a binary system, pulsar spindown, etc. and it is therefore only the poloidal component which we observe directly. The magnetic field {\it inside} the star, including the toroidal component, can be important however in other ways, for instance as an energy source for magnetars and as a means of creating `hot spots' on neutron stars via anisotropic heat conduction. Also, the Lorentz force gives rise to a distortion of the density field which has two main consequences: firstly, by changing the moment of inertia and making the star triaxial, free precession is expected. There is observational evidence of precession in some stars \citep{Akguenetal:2006}. Damping of this free precession results in a star with its rotation and magnetic axes either parallel or perpendicular, depending on whether the star is oblate or prolate, respectively. A spinning neutron star with non-parallel axes should be observable via gravitational radiation {\mk (see e.g. \citealt{DalandSte:2007} and references therein), especially a fast-spinning magnetar as required by `millisecond magnetar' models of SNe.} An important question is therefore whether the field inside the star is dominated by the poloidal or by the toroidal components, since the former makes the star oblate and the latter prolate, as well as because the decay of a large `hidden' toroidal field can provide an energy source for magnetars.

The energy argument of Section~\ref{sec:energy} as applied to axisymmetric fields gives us an important clue as to the permissible toroidal/poloidal ratios in a stable axisymmetric equilibrium: the poloidal field cannot be much stronger than the toroidal, or it would be energetically favourable to make a transition to non-axisymmetric equilibrium. However, it cannot be ruled out that the poloidal field is still strong enough that the star is oblate; more quantitative calculations are required. It also cannot be ruled out that the toroidal field can be much stronger than the poloidal; in this situation a non-axisymmetric equilibrium would adjust, the flux tubes becoming shorter and fatter, but an axisymmetric flux tube cannot become any shorter and fatter ($\alpha$ cannot go above $180^\circ$). The question of exactly how much stronger the toroidal field can be has to do with an instability of the toroidal field itself \citep{Tayler:1973}; this will be examined in a forthcoming paper.

\section{Conclusions}\label{sec:conclusion}

Numerical magnetohydrodynamic simulations of a stably-stratified star have been performed with an initially random magnetic field. It is found that this initial magnetic field always evolves on the Alfv\'en time-scale {\mk into a stable equilibrium configuration consisting of twisted flux tube(s). There are approximately axisymmetric equilibria with one flux tube forming a circle around the equator, and more complex, non-axisymmetric equilibria consisting of one or more flux tubes arranged in a more complex pattern with their axes lying at roughly constant depth under the surface of the star. Whether an axisymmetric or non-axisymmetric equilibrium forms depends on the radial profile of the initial field strength: a more centrally concentrated field evolves into an axisymmetric equilibrium and a more spread-out field with greater flux connection to the atmosphere evolves into a more complex equilibrium\footnote{{\mk A magnetised region of initially `turbulent' field, embedded in an initially stationary homogenous conducting gas, is found to evolve into an axisymmetric equilibrium (this will be explored in a forthcoming paper.) It seems then that {\it non}-axisymmetric equilibria form only because of the non-conducting nature of the atmosphere.}}. Further, higher-resolution simulations are required to quantify better this distinction in initial conditions, but it seems that} in an ideal-gas star with polytropic index $n=3$, if the initial field strength is tapered as $B\sim\rho^p$ then the threshold is $p\sim1/2$.


{\mk These equilibria evolve quasi-statically due to diffusive processes (finite conductivity, thermal/composition diffusion). Axisymmetric equilibria rise towards the surface and eventually make a transition to the non-axisymmetric class; the non-axisymmetric configurations undergo a gradual lengthening and narrowing of their flux tubes. This is caused by loss of toroidal flux into the atmosphere.
In configurations with more than one flux tube,} each tube may have either positive or negative magnetic helicity; whether negative or positive has no effect on the stability. Stable zero-helicity equilibria are possible.



{\it Acknowledgments.} The author would like to thank Jonathan Dursi, \AA ke Nordlund, Andreas Reisenegger, Henk Spruit, Chris Thompson, Gregg Wade and Anna Watts for fruitful discussions and suggestions. Some of the figures were made with {\sc vapor} ({\tt www.vapor.ucar.edu}).

\bsp

\end{document}